\documentclass{iopart}[12pt]
\usepackage{iopams}
\usepackage{epsfig}
\usepackage{times}
\usepackage{pstricks}
\usepackage{amsfonts,amsthm,eucal,amssymb}
\newcommand{\sn}{\mbox{sn}}
\newcommand{\cn}{\mbox{cn}}
\newcommand{\dn}{\mbox{dn}}
\newcommand{\re}{\mbox{Re}}
\newcommand{\im}{\mbox{Im}}
\newcommand{\am}{\mbox{am}}

\eqnobysec

\begin{document}

\jl{1}

\title{A New Effective Mass Hamiltonian and Associated Lam\'e Equation: Bound States}

\author{A.~Ganguly$^{1,2}$\footnote{e-mail:
gangulyasish@rediffmail.com}, M.~V.~Ioffe$^{2,3}$\footnote{e-mail:
m.ioffe@pobox.spbu.ru} and L.~M.~Nieto$^{2}$\footnote{e-mail:
luismi@metodos.fam.cie.uva.es}}

\address{$^1$\ City College, University of Calcutta, 13 Surya
Sen Street, Kolkata -- 700012, India}

\address{$^2$\ Departamento de F\'{\i}sica Te\'orica, At\'omica y \'Optica,
Universidad de Valladolid, 47071 Valladolid, Spain}

\address{$^3$\ Department of Theoretical Physics, Sankt-Petersburg
State University, 198504 Sankt-Petersburg, Russia}

\date{\today}

\begin{abstract}
A new quantum model with rational functions for the potential and
effective mass is proposed in a stretchable region outside which
both are constant. Starting from a generalized effective mass
kinetic energy operator the matching and boundary conditions for
the envelope wave functions are derived. It is shown that in a
mapping to an auxiliary constant-mass Schr\"odinger picture one
obtains one-period ``associated Lam\'e'' well bounded by two
$\delta$-wells or $\delta$-barriers depending on the values of the
ordering parameter $\beta$. The results for bound states of this
new solvable model are provided for a wide variation of the
parameters.
\end{abstract}


 \vspace{.5cm}
 \noindent
 \textit{Accepted in J. Phys. A}

\section{Introduction}

Effective mass theory had been used for years in several
 branches of modern physics like nuclear physics \cite{pre} or
  solid-state physics (see e.g. the pioneering works \cite{wan}--\cite{lut}).
  This theory is an useful tool for studying the motion
of carriers in pure crystals and also for the virtual-crystal
approximation to the treatment of homogeneous alloys (where the
actual one-electron potential is approximated by a periodic
potential), as well as of graded mixed semiconductors (where
virtual-crystal potential is not periodic). The salient feature of
this theory is that it approximates a complicated physical
situation to the solution of a Schr\"odinger equation with
position-dependent effective mass function, the so-called
effective mass (EM) eigenvalue equation.  The position-dependent
EM is also used in the construction of pseudo-potentials, which
have a significant computational advantage in quantum Monte Carlo
method \cite{fou}. Needless to emphasize that the growing interest
in semiconductor physics parallel to the modern development in
fabricating nanostructure technology creates a renewed attention
to study the behavior of one-dimensional EM eigenvalue equation
\begin{equation}\label{eigen1}
H_{EM}(x)\psi (x)\equiv [T_{EM}(x)+V(x)]\psi (x)=E\psi (x)
\end{equation}
from a theoretical standpoint. In this equation
 $T_{EM}(x)$ is the EM kinetic energy operator and $\psi(x)$ is the EM wave
function (also known as `envelope wave function' in the literature
 \cite{lut}--\cite{pis}).
 It has to be mentioned that the justification of this EM
approximation in the context of a realistic situation is not the
purpose of the present study, rather we will concentrate on the
analysis of the spectral properties of the EM eigenvalue equation
(\ref{eigen1}). The first step is certainly to choose a suitable
form of the Hermitian kinetic energy operator arising from
non-commutativity of momentum operator $p\equiv -i\hbar\partial_x$
and the effective mass operator $m(x)$. Different forms had been
proposed in the literature, most of which may be written as a
special class of the general two-parameter family proposed in
Ref.~\cite{von}
\begin{equation}\label{kin}
T_{EM}(x)=\frac{1}{4} \left(m^{\alpha}pm^{\beta}pm^{\gamma}+m^{\gamma}pm^{\beta}pm^{\alpha}\right)
\end{equation}
with the constraint $\alpha+\beta+\gamma=-1$ over the ordering
parameters.

Considerable efforts were made to remove the non-uniqueness of the
kinetic energy operator (\ref{kin}) or, in other words, to fix the
values of the ordering parameters $\alpha,\beta,\gamma$. In
\cite{mor1} a step potential and a step mass were considered and
it was shown that $\alpha=\gamma$ is the only physical choice for
an abrupt heterojunction. Later in an attempt to fix $\beta$, two
different conclusions were drawn, namely $\beta=0$ for a
one-dimensional model \cite{mor2} and $\beta=-1$ for a
three-dimensional model \cite{mor3}. On the other hand, in a
series of works \cite{ein1,ein3,ein4}, the authors concluded that
$\alpha=\gamma=0,\beta=-1$ for an abrupt heterojunction. Among
these works, Ref.~\cite{ein4} deserves to be mentioned separately
because it presented the first example of a continuous function
$m(x)$ across the heterojunction. A new kind of kinetic energy
operator was proposed \cite{ein2} for strained heterostructure,
which is not included in (\ref{kin}), in general, for
position-dependent lattice constant. It should be mentioned that
the choice $\alpha=\gamma=0,\beta=-1$ gives rise to the kinetic
energy operator $T_{EM}=p(1/2m)p$, which was first proposed in
Ref.~\cite{ben}. Choosing the same operator some interesting
pedagogical models were considered \cite{lev1,lev2} to show the
qualitative differences in quantum mecanical observables (e.g.
reflection and transmission coefficient, band-structure, etc.)
between EM and constant-mass case. Many other forms of kinetic
energy operator had also been proposed, e.g., $\alpha=-1$,
$\beta=0$, and $\gamma=0$ which gives from (\ref{kin})
$T_{EM}=(1/4m)p^2+p^2(1/4m)$ \cite{gora};
$\alpha=\gamma=-1/2,\beta=0$ which yields $T_{EM}=(1/2\sqrt{
m})p^2(1/2\sqrt{m})$ \cite{zhu}, etc. A different variation was
derived, via path-integral formalism \cite{yung}, which comes from
(\ref{kin}) for the following values:
 $\alpha=(-\sqrt{2}+i)/3\sqrt{2} ,\beta=-1/3 ,\gamma=\alpha^*$.
 It is therefore clear that no universal choice for the ordering parameters exists
in the literature of EM theory.

In recent times several authors [28--50] either started from a
preferred ad-hoc choice for $\alpha,\beta,\gamma$ or kept them
arbitrary. In both cases attention had been paid to solvability
for various smooth functional forms of $V(x)$ and $m(x)$ by
employing the existing tools like supersymmetry [29, 30, 34, 35,
40, 44--48], Lie-algebraic approach \cite{roy,koc1,koc2,bag2},
shape-invariance \cite{koc4,bag4}, etc.  The connection between
solvability and the ordering parameters in equations
(\ref{eigen1}) and (\ref{kin}) was discussed in Ref.~\cite{dut}.
On the full line, smooth functions (in the sense that $m'$ and
$m''$ are also continuous) were chosen for the first time in
\cite{dek}, where the authors however concluded again that
$\alpha=\gamma=0,\beta=-1$ by comparing their results with a
limiting case where the potential
 and mass become abrupt. It may be mentioned that in several works [33--35] 
mass function was kept arbitrary
and thus the solutions provided there were only formal. On the
contrary, in most cases where smooth functional forms were chosen
for $m(x)$, we notice that $m(x)\rightarrow 0$ as $|x|\rightarrow
\infty$. One possible way of eliminating this nonphysical
situation is to consider the variation of the mass in a finite
region (being constant outside), which is also natural on
realistic ground. Of course, this correction will force us to
obtain correct matching conditions for the model.

Quite justifiably we will use the generic kinetic energy operator
(\ref{kin}) without imposing any additional constraint over the
ordering parameters. One of the purpose of our present work is to
show that a sensible quantum situation could be modelled by a
continuous functional form for $V(x)$ and $m(x)$, which are free
from the above mentioned defect of vanishing mass at infinity. A
new combination of rational forms  will be chosen for both
functions inside a region, but outside they will be considered to
be constant, that is to say, our EM Hamiltonian is asymptotically
equivalent to conventional constant-mass Hamiltonian. Our other
purpose is to study the properties of bound states of this model
and to compare the results with the constant-mass problem. The
precise model chosen in the present work for $m(x)$ and for the
potential is motivated by the following property: it allows the
transformation to the auxiliary spectral problem of a conventional
(constant mass) Schr\"odinger equation, which is solvable. In this
sense this variable mass problem can also be characterized as
solvable.

The structure of the paper is as follows. In Sec.~2 we will
introduce our model and derive the appropriate matching
conditions. Applying these conditions the equation for the
energies of the bound-states will be obtained (the details of some
of the issues dealt in this section will be provided in the
 \ref{appendix}, for readers' convenience). In Sec.~3, we
will map the whole problem in an auxiliary constant-mass
Schr\"odinger picture and obtain the correct physical range of
bound-state energies of our EM potential. A limiting case will be
considered in Sec.~4, where the potential becomes the well-known
harmonic oscillator. In Sec.~5, we will solve numerically the
transcendental energy equation for bound states and examine the
spectral properties for a wide range of the parameters. Some
characteristic differences from the constant-mass case will be
noted here. Sec.~6 will contain a description of some interesting
features of bound-state wave functions, obtained by the analysis
of the numerical results of the previous section. Finally, we will
end with our conclusion in Sec.~7.
\begin{figure}
\centering
\begin{pspicture}(-2.5,-1)(3,2)
\psset{xunit=1.2cm}\psset{yunit=1.0cm}
\pscurve[linewidth=1.5pt]{-}(-4,0.18)(-2,0.18)(-1.78,0.18)
\pscurve[linewidth=1.5pt]{-}(-1.78,0.18)(-1.4,0.28)(-0.5,0.78)(0,1)(0.5,0.78)(1.4,0.28)(1.78,0.18)
\pscurve[linewidth=1.5pt]{-}(1.78,0.18)(2,0.18)(4,0.18)
\pscurve{<->}(-4.5,-1)(0,-1)(4.9,-1)
\psline{-}(-1.78,-1)(-1.78,2)

\psline{-}(1.78,-1)(1.78,2)
\pspolygon[linestyle=none,fillstyle=vlines,hatchsep=8pt](-1.78,2)(-1.78,-1)(1.78,-1)(1.78,2)%
\pspolygon[linestyle=none,fillstyle=hlines,hatchsep=8pt](-4,2)(-4,-1)(-1.78,-1)(-1.78,2)%
\pspolygon[linestyle=none,fillstyle=hlines,hatchsep=8pt](4,2)(4,-1)(1.78,-1)(1.78,2)%
\rput(3,1.4){\psframebox*[framearc=.3]{Material 1}}%
\rput(-3,1.4){\psframebox*[framearc=.3]{Material 1}}%
\rput(0,1.4){\psframebox*[framearc=.3]{Material 2}}%
\pscurve[linewidth=1.5pt]{-}(1.78,-0.36)(2,-0.36)(4,-0.36)
\pscurve[linewidth=1.5pt]{-}(-1.78,-0.36)(-1.4,-0.16)(-1.2,-0.19)(-0.5,-0.47)(0,-0.775)(0.5,-0.47)(1.2,-0.19)(1.4,-0.16)(1.78,-0.36)
\pscurve[linewidth=1.5pt]{-}(-1.78,-0.36)(-2,-0.36)(-4,-0.36)
\rput(-1.78,-1.3){\footnotesize{$-x_0$}}
\rput(1.78,-1.3){\footnotesize{$x_0$}}
\rput(5.0,-1){\footnotesize{$x$}}%
\rput(4.7,0.18){\footnotesize{$\leftarrow m(x)$}}%
\rput(4.7,-0.36){\footnotesize{$\leftarrow V(x)$}}%
\rput(-4.3,0.18){\footnotesize{$m_0$}}%
\rput(-4.3,-0.36){\footnotesize{$V_0$}}%
\end{pspicture}
\caption{A schematic representation of the model: a variable mass
and a potential well in the interval $(-x_0,x_0)$, and outside of
this interval both are constant.} \label{fig1}
\end{figure}
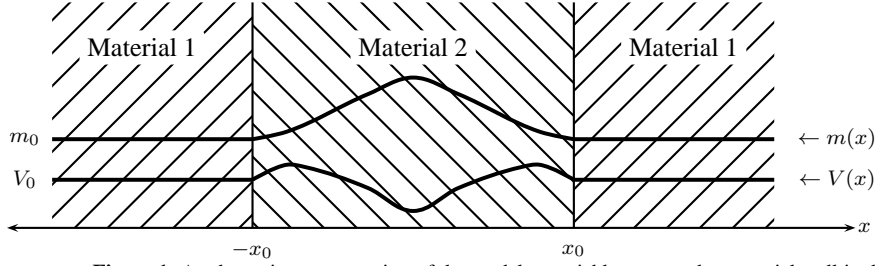

\section{A model Hamiltonian and the energy equation}\label{model}

\subsection{The model}

To start with, we will introduce a hypothetical sample shown in
Fig.~\ref{fig1}, consisting of two materials which produces a
variation of mass in a stretchable region $(-x_0,x_0)$. We are
interested in this exposition to study the bound states of a
particle in the presence of a local potential $V(x)$
\begin{equation}\label{emp}
V(x) = \left\{
\begin{array}{ll}
 f_{V}(x), &  |x|\leq x_0,  \\ [1ex] f_V(x_0)\equiv V_0=\mbox{constant}, & |x|>x_0
\end{array} \right.
\end{equation}
where the function $f_V(x)$ is chosen to be
\begin{equation}
 f_V(x)=\frac{Ak^2}{1+x^2}+\frac{Bk^2}{1+k'^2x^2}+Ck'^2x^2+D.
\end{equation}

The potential function depends on three classes of parameters : i)
the ordering parameters $\alpha,\beta,\gamma\in \mathbb{R}$ of the
kinetic term (\ref{kin}), such that $\alpha+\beta+\gamma=-1$ , ii)
the elliptic modulus parameter $k^2\in (0,1)$ or the complementary
modulus $k'^2=1-k^2$, and iii) the ``Lam\'e parameters'' $\mu,\nu
\in \mathbb{N}$ (this terminology will be explained shortly). The
four constants $A,B,C,D$ are expressed in terms of these
parameters as
\begin{equation} \label{def1}
\hspace{-1cm}\begin{array}{ll}
 A  =  4(1+\beta-\eta)-\left (\mu+\frac{1}{2}\right )^2, &
 B  =  \left (\nu+\frac{1}{2}\right )^2-4(1+\beta-\eta), \\ [1ex]
 C  =  2 \, [\, 8\eta-5\beta-6\, ]-2\, , &
 D  =  (1+\beta)(3k^2+2)-4\eta k^2+\frac{k^2}{4}-1.
\end{array}
\end{equation}
In the above expressions the quantity $\eta$ stands for $\eta  =
1+\beta+\alpha(\alpha+\beta+1)$. %
The point $x_0$ depends on $k$ by $x_0=1/\sqrt{k'}$ and the mass
function is given by
\begin{equation}\label{em}
m(x) = \left\{
\begin{array}{ll}
 f_m(x), & \  | x|\leq x_0, \\ [1ex] f_m(x_0)\equiv m_0=\mbox{constant}, & \ |x|>x_0
\end{array} \right.
\end{equation}
where
\begin{equation}\label{def2}
f_m(x)=\left [(1+x^2)(1+k'^2x^2)\right ]^{-1}.
\end{equation}

It should be stressed that the choice of this particular model is
made because it not only provides a continuous mass-function with
physically reasonable non-vanishing limit at infinity, but also
leads to well-known associated Lam\'e equation, as we shall show
subsequently. Thus, our task now is to solve the eigenvalue
equation (\ref{eigen1}) for the general kinetic energy operator
(\ref{kin}) with the potential $V(x)$ and mass $m(x)$ given by
(\ref{emp})--(\ref{def2}). Choosing for our convenience the scale
$\hbar^2=2$, this equation may be written as
\begin{equation}\label{eigen2}
 \hspace{-1.3cm}\psi''(x)-\frac{m'}{m}\psi'(x)+\left [m\{E-V(x)\}-\frac{1+\beta}{2}\frac{m''}{m}+\eta
\left( \frac{m'}{m}\right)^2 \right ]\psi(x)=0,
\end{equation}
where throughout this article prime and dot will denote
derivatives with respect to $x$ and $z$, respectively.

\subsection{Matching conditions}\label{matching}
To derive the correct matching and boundary conditions for the
envelope wave function $\psi(x)$, we notice that the presence of
$m''/m$ in (\ref{eigen2}) produces a $\delta$-discontinuity at the
two junctions $x=\pm x_0$. Thus, $\psi'$ must have a definite jump
at these junctions to balance these singularities in the EM
eigenvalue equation (\ref{eigen2}). To calculate precisely these
jumps, we will express $m(x)$ and $V(x)$ in terms of the Heaviside
$\Theta$-function
\begin{equation}
 \begin{array}{ll}
\Theta (x) = \left\{
\begin{array}{ll}
 1, & x>0, \\ [1ex]
1/2, & x=0, \\ [1ex]
0, & x<0,
\end{array}
\right. & \qquad [\Theta'(x)=\delta(x)]
 \end{array}
\end{equation}
as
\begin{equation}\label{emtheta}
 \hspace{-1.2cm}m(x)  =  m_0 \, \Xi(x)+f_m(x)\, [1-\Xi(x)]\, , \qquad
V(x)  =  V_0 \, \Xi(x)+f_V(x)\, [1-\Xi(x)],
\end{equation}
 where
\begin{equation}
 \Xi(x)  =  \Theta (-x_+)+\Theta (x_-), \quad x_{\pm}  =  x\pm x_0.
  \label{deft}
\end{equation}

It may be noticed at once that, due to cancellation effect, $m'$
will not contain $\delta$-discontinuity.\label{cancellation} Thus,
we get
\begin{equation}\label{empt}
 \hspace{-1.2cm}m'(x)  =  f'_m(x)\, [1-\Xi(x)]\, , \qquad
m''(x)  =  f''_m(x) \, [1-\Xi(x)]-f'_m(x) \, \Xi'(x).
\end{equation}
It is obvious that the first order derivative term in (\ref{eigen2})
could be eliminated by introducing the transformed function
\begin{equation}\label{tr1}
\phi(x)=\psi(x)/\sqrt{m(x)}\ .
\end{equation}
The EM eigenvalue equation (\ref{eigen2}) reduces to
\begin{equation}\label{eigen2a}
 \phi''(x)= \left[m\{V(x)-E\}+\frac{\beta}{2}\frac{m''}{m}+\left(\frac{3}{4}-\eta\right)
       \left(\frac{m'}{m} \right)^2 \right]\phi(x)\, .
\end{equation}
 Integration of equation (\ref{eigen2a}) with respect to $x$ in
 the interval $[x_0-\epsilon,x_0+\epsilon]$ yields
\begin{eqnarray}
&&
 \hspace{-1.5cm}\phi'(x) \Bigr|^{x_0+\epsilon}_{x_0-\epsilon}=\int^{x_0}_{x_0-\epsilon}
 \left [ f_m \{f_V-E\}+\frac{\beta}{2}\frac{f''_m}{f_m}
  +\left (\frac{3}{4}-\eta\right )\left (\frac{f'_m}{f_m}\right )^2 \right ]\phi(x) \nonumber \\ [1ex] &&
 \hspace{.5cm}\mbox{}+m_0(V_0-E)\int^{x_0+\epsilon}_{x_0}\phi(x)dx
   -\frac{\beta}{2}\int^{x_0+\epsilon}_{x_0-\epsilon}\frac{f'_m}{f_m}
   \delta(x_-)\phi(x)dx. \label{int}
\end{eqnarray}
Let us notice that for our model the wave function $\psi(x)$ must
be continuous\footnote{The case of discontinuity in both $V(x)$
and $m(x)$ (e.g. potential-mass step \cite{mor1} or a quantum well
with mass mismatch) corresponds to discontinuity of $\psi(x)$
\cite{as9}.} on the full line, for otherwise $\psi''(x)$ would
have stronger singularity than that from the term $m''$ in
equation (\ref{eigen2}). It then follows from (\ref{tr1}) that
$\phi(x)$ is a continuous function. Hence all integrands in
(\ref{int}) are
 continuous except the last term. Thus letting $\epsilon\rightarrow 0$,
 we will obtain the jump at $x=x_0$:
\begin{equation}
 \left . \triangle \phi'\right |_{x=x_0}=-\frac{\beta}{2}\frac{f'_m(x_0)}{f_m(x_0)}\phi (x_0).
\end{equation}
Proceeding similarly for the other junction $x=-x_0$:
\begin{equation}
 \left . \triangle \phi'\right |_{x=-x_0}=\frac{\beta}{2}\frac{f'_m(-x_0)}{f_m(-x_0)}\phi (-x_0).
\end{equation}

We thus derive the following conditions for the envelope wave
function $\psi$ and its derivative
\begin{displaymath}
   \mbox{a) $\psi(x)$ is continuous} \, ,
\end{displaymath}

\vspace*{-.3cm}
\begin{equation}\label{dc}
\hspace{-.069cm}\mbox{b) } \left .\triangle \left
(\frac{\psi}{\sqrt{m}} \right )'
  \right |_{x=\pm x_0}\!\!=\mp \frac{\beta}{2}\frac{f'_m(\pm x_0)}{f_m (\pm
   x_0)}\frac{\psi(\pm x_0)}{\sqrt{m(\pm x_0)}}\, ,
\end{equation}
and for the bound state wave functions
\begin{equation}
  \label{nc}\hspace{-.04cm}\mbox{c)}\int_{-\infty}^{\infty}|\psi_n(x)|^2 dx=1, \quad (\mbox{normalizability}).
\end{equation}

\subsection{Expressions for wave functions}
The EM eigenvalue equation (\ref{eigen2a}) takes the following form
in the three regions
\begin{equation}\label{eigen3}
 \hspace{-1.5cm} \phi''(x)  = \left\{
\begin{array}{ll}
 \kappa^{2}\phi(x), & |x|>x_0 \\ [1ex]
 \left[ f_m(x)\left \{f_V(x)-E \right \}+\frac{\beta}{2}\frac{m''}{m}
     + \left (\frac{3}{4}-\eta \right )
                 \left (\frac{m'}{m} \right )^2 \right]\phi(x), & |x|<x_0,
\end{array}
\right.
\end{equation}
in which we have used the abbreviation
\begin{equation}
\label{def3}\kappa^2=m_0(V_0-E).
\end{equation}

In this exposition we are primarily interested in bound states and
thus the quantity $\kappa$ in (\ref{def3}) will be always real and
nonzero. The scattering states for which $E\geq V_0$ will be
discussed at length elsewhere. The acceptable solutions of
(\ref{eigen3}) for $|x|>x_0$ are
\begin{equation}
\label{w1}
\phi(x) = \left\{
\begin{array}{ll}
\mathcal{N}^- e^{\kappa x}, & x<-x_0 \\ [1ex]
\mathcal{N}^+ e^{-\kappa x}, & x>x_0,
\end{array}
\right.
\end{equation}
where $\kappa$ is taken as positive square root of (\ref{def3})
and the constants $\mathcal{N}^{\pm}$ have to be determined from
the conditions a)--c). In the intermediate region the
reformulation of equation (\ref{eigen3}) is useful to solve the
spectral problem. The crucial observation is that this equation
may be transformed into the well-known associated Lam\'e equation
[52--57]
\begin{equation}
 \label{lame}
\ddot\chi(z)-k^2 \left
[\mu(\mu+1)\sn^2z+\nu(\nu+1)\frac{\cn^2z}{\dn^2z} -\frac{E}{k^2}
\right ]\chi(z)=0
\end{equation}
by means of the following changes of variables
\begin{equation}\label{tri}
 x(z)=\sn z/\cn z, \qquad \chi(z)=\left [f_m(x(z))\right ]^{1/4}\phi(x(z))\, .
\end{equation}
In the above equations $\sn z\equiv \sn(z,k),\cn z\equiv \cn(z,k)$
and $\dn z\equiv\dn(z,k)$ are three Jacobian elliptic functions of
real modulus $k$. Since the parameters $\mu,\nu$ enter in the
associated Lam\'e potential (\ref{lame}), we will reserve the
terminology ``Lam\'e parameters'' for them. Note that the equation
(\ref{lame}) is of period $K$ or $2K$ according as $\mu=\nu$ or
$\mu\neq\nu$, where
$K(k)=\int_0^{\pi/2}d\tau/\sqrt{1-k^2\sin^2\tau}$ is the complete
elliptic integral of second kind. The Lam\'e parameters can be
chosen as any integral pair, but for simplicity we will choose
$\mu=\nu=1$. In this context it may be emphasized that the
associated Lam\'e equation (\ref{lame}) for $\mu=\nu$ can be
mapped via coordinate transformation $\tilde{z}=(1+k')z$ to
ordinary Lam\'e equation with a different modulus parameter
$\tilde{k}=(1-k')/(1+k')$ and the energy variable
$\tilde{E}=-\mu(\mu+1)\tilde{k}+E/(1+k')^2$ through the use of the
relation
$\sn(\tilde{z},\tilde{k})=(1+k')\sn(z,k)\cn(z,k)/\dn(z,k)$.
However in our auxiliary variable $z$, equation (\ref{lame})
represents associated Lam\'e equation, and for $\mu=\nu$ it
becomes $K$-periodic. This means that we are going to consider
this equation in a single period $(-K/2,K/2)$.

At this stage it is worth mentioning that  the region $(-x_0,x_0)$
of material 2 must be so chosen that it may be stretched as large
as we please, but it will be always finite (it must be surrounded
by material 1). On the other hand, it can not be shrunk to a point
due to the presence of material 2 (see Fig.~\ref{fig1}). For this
reason we need to exclude the points $z=\pm K$ from the domain of
equation (\ref{lame}), as it can be easily verified from the
transformation (\ref{tri}) that these points lead to
$x=\pm\infty$. For definiteness we have chosen the interval $z\in
(-K/2,K/2)$ which just corresponds to $x\in(-x_0,x_0)$ where
$x_0=1/\sqrt{k'}$. Let us remark about two well-known limits
$k\rightarrow 1$ and $0$, which are usually considered for
elliptic functions. Note that as $k\rightarrow 1$, the
complementary modulus $k'\rightarrow 0$ and consequently $x_0$
goes to infinity. Thus $k\rightarrow 1$ limit is not allowed in
our model. But the other limit $k\rightarrow 0$ may be considered
as it allows the shrinking of the region $(-x_0,x_0)$ up to a
finite interval $(-1,1)$. The general solutions of equation
(\ref{lame}) for arbitrary energy $E$, which we need, was obtained
only recently in Ref \cite{as7,as8}. Here we will not describe the
method of obtaining these solutions, but for readers' convenience,
we have included a self-contained brief introduction about
elliptic functions in \ref{appendix}. The two linearly independent
solutions\footnote{The notations for the Lam\'e parameters
$\mu,\nu$ in Ref.~\cite{as7,as8} are $m,\ell$.} of (\ref{lame})
are (for three exceptional cases see below)
\begin{equation}\label{ls}
 \hspace{-1.5cm}\chi_{1,2}(z)=\frac{\prod_{i=1}^{2}\sigma (z-iK'\pm a_i)}{\sigma(z-iK'+\omega_1)\sigma(z-iK')}
          \exp \left [(z-iK')\{\zeta (\omega_1)\mp\zeta(a_1)\mp\zeta (a_2)\}\right ],
\end{equation}
where $a_i(E)$ are to be determined from the equation
$\wp(a_i)=c_i$, $c_i$ being zeros of the following quadratic
equation (see Eqs. (4)--(5) in Ref \cite{as7})
\begin{equation}\label{cc}
  c^2+(E-4+e_1)c+(3e_1-e_2-2e_1e_3-e_1E)=0\, .
\end{equation}
Here, $e_i$ are always real $(e_1>e_2>e_3)$ defined by
$\wp(\omega_i)=e_i$; $\omega_1$ and $\omega_3$ are half-periods of
Weierstrass elliptic function $\wp(z)$ (see  \ref{appendix} for
more details). For the numerical convenience, here we have chosen
the scale $e_1-e_3=1$ so that $\omega_1=K, \, \omega_3=iK', \,
\omega_2=\omega_1+\omega_3, \: K'(k)=K(k')$. Thus, the solution in
the intermediate region takes the form
\begin{equation}\label{ifs}
  \chi(z)=d_1\chi_1(z)+d_2\chi_2(z)\, .
\end{equation}
The constants $d_1,d_2$ have to be fixed from the  conditions
a)--c). Before proceeding to do that, we would like to point out
that for special values of $E=E^{(j)},j=0,1,2$ both solutions
$\chi_1(z)$ and $\chi_2(z)$ become identical and coincide to
periodic (or anti-periodic) band-edge solutions $\chi^{(j)}$. The
explicit expressions for these solutions \cite{as3} and
corresponding values of $a_i$ in (\ref{cc}) are
\begin{equation} \label{bs}
 \hspace{-1.5cm}\begin{array}{lll}
\chi^{(0)}=\dn z+k'/\dn z , & E^{(0)}=2+k^2-2k', &
              a_1=-a_2=\omega_1/2\, , \\
 \chi^{(1)}=\dn z-k'/\dn z , & E^{(1)}=2+k^2+2k', &
              a_1=-a_2=-\omega_3+\omega_1/2\, , \\
\chi^{(2)}=\sn z \, \cn z/\dn z , & E^{(2)}=4, &
              a_1=\omega_3,\quad a_2=\omega_2\, .
  \end{array}
\end{equation}
Thus, for $E=E^{(j)}$ the two linearly independent solutions of
associated Lam\'e equation (\ref{lame}) may be given as
\begin{equation}\label{soldeg}
  \chi_1(z)=\chi^{(j)}(z), \quad
  \chi_2(z)=\chi_1(z)\int^z\frac{d\tau}{[\chi_1(\tau)]^2}\, .
\end{equation}

It should be emphasized that in our model the situation is
different from that of periodic associated Lam\'e equation
\cite{as3,as4} because the equation (\ref{lame}) is considered in
only one period $(-K/2,K/2)$. We will show later that the discrete
energy levels of our EM Hamiltonian lie inside the allowed bands
of constant-mass periodic associated Lam\'e Hamiltonian, and
corresponding wave functions $\psi(x)$ are obtained from
(\ref{w1}) and (\ref{ls}).

\subsection{Energy equation for bound states}
The continuity condition and slope-discontinuity requirement
(\ref{dc}) can be expressed in a single pair of equations
\begin{equation} \label{ca}
 \hspace{-1cm}\mathcal{N}^{\pm}m_0^{\frac{1}{4}}e^{-\kappa/\sqrt{k'}}=d_1\chi_1^{\pm}+d_2\chi_2^{\pm},
\end{equation}

\vspace*{-.3cm}
 \noindent
\begin{equation}\label{da}
 \hspace{-2.2cm}\mathcal{N}^{\pm}(\beta\sqrt{k'}+\kappa)e^{-\kappa/\sqrt{k'}}=
    d_1 ( \mp m_0^{\frac{1}{4}}\dot \chi_1^{\pm}-\frac{\sqrt{k'}}{2}m_0^{-\frac{1}{4}}\chi_1^{\pm} )
              +d_2 (  \mp m_0^{\frac{1}{4}}\dot \chi_2^{\pm}-\frac{\sqrt{k'}}{2}m_0^{-\frac{1}{4}}\chi_2^{\pm} )
\end{equation}
where we have used the abbreviations
\begin{equation}\label{def4}
 \chi_i^{\pm}=\chi_i(\pm K/2), \quad \dot \chi_i^{\pm}=\dot\chi_i(\pm K/2), \: i=1,2.
\end{equation}
Eliminating $\mathcal{N}^{\pm}$ from (\ref{ca})--(\ref{da}), we
obtain a homogeneous linear system for $d_1,d_2$
\begin{eqnarray*}
 d_1[ 2m_0^{1/4}\dot \chi_1^++m_0^{-1/4}\mathcal{B}\chi^+_1]
           +d_2[2m_0^{1/4}\dot\chi_2^++m_0^{-1/4}\mathcal{B}\chi^+_2]=0, \\ [1ex]
   d_1[2m_0^{1/4}\dot \chi_1^--m_0^{-1/4}\mathcal{B}\chi^-_1]
         +d_2[2m_0^{1/4}\dot\chi_2^--m_0^{-1/4}\mathcal{B}\chi^-_2]=0,
\end{eqnarray*}
where the quantity $\mathcal{B}$ reads
\begin{equation}\label{def5}
\mathcal{B}=(2\beta+1)\sqrt{k'}+2\kappa \, .
\end{equation}
Demanding for non-trivial
solutions of $d_1,d_2$ from the previous system of equations, we have obtained our energy equation in the following
form
\begin{equation}\label{enp}
  4 m_0 (T^-_2-T^+_2)+2\sqrt{m_0}\mathcal{B}
  (\dot{T}^-_1+\dot{T}^+_-)+\mathcal{B}^2(T^+_1-T^-_1)=0,
\end{equation}
where
\begin{equation}\label{def1a}
  T^{\pm}_1=\chi_1^{\pm}\chi_2^{\mp};\quad
  \dot{T}_1^{\pm}=\dot{\chi}_1^{\pm}\chi_2^{\mp}-\chi_1^{\pm}\dot{\chi}_2^{\mp};\quad
  T_2^{\pm}=\dot{\chi}_1^{\pm}\dot{\chi}_2^{\mp}.
\end{equation}
In particular for $E\neq E^{(j)}, j=0,1,2$, the energy equation
(\ref{enp}) may be further simplified by inserting the expressions
for $\dot{\chi}_i^{\pm}$ from equation (\ref{ls}) (see
\ref{appendix}) and then the energy equation reduces to the form
\begin{equation}\label{en}
 \hspace{-1.8cm}\left [ 4m_0A_-^2-4\mathcal{B}\sqrt{m_0}A_-+\mathcal{B}^2\right ]T_1^-
   -\left [4m_0A_+^2+4\mathcal{B}\sqrt{m_0} A_++\mathcal{B}^2\right ]T_1^+=0,
\end{equation}
where $A_{\pm}$ are given by
\begin{equation}\label{defa}
A_{\pm}=\pm\frac{1}{2}\left [\sum_{i=1}^{2}\frac{\dot\wp\left
(\frac{\omega_1}{2}+\omega_3\right )\mp\dot\wp(a_i)}
        {\wp\left (\frac{\omega_1}{2}+\omega_3\right )-\wp(a_i)}-\frac{\dot\wp\left (\frac{\omega_1}{2}+\omega_3\right )}
        {\wp\left (\frac{\omega_1}{2}+\omega_3\right )-e_1} \right ]
\end{equation}

Hence, similar to the well-known textbook example of the finite
square well potential, we have to deal with the transcendental
energy equations (\ref{enp}) and (\ref{en}). It is clear from the
structure of equation (\ref{en}) that for $T^-_1=T^+_1$, some
roots will be given by the equation $A_--A_+=0$. But one may
notice from (\ref{defa}) that this equation will be true if and
only if $\sum \dot\wp (a_i)=0$. The latter condition is, however
satisfied only for band-edge energies $E=E^{(j)}$, mentioned in
the equation (\ref{bs}). This clearly implies that the roots
$E=E^{(j)}$ of (\ref{en}) have to be discarded, because for these
values the correct energy equation is (\ref{enp}), where $\chi_i$s
are given by (\ref{bs}) and (\ref{soldeg}). It may be mentioned
that we have checked numerically for different values of the
potential parameters that the energy equation (\ref{enp}) is not
satisfied for $E=E^{(j)}$. This means that for those cases the
band-edge energies of periodic associated Lam\'e Hamiltonian are
not roots of the energy equation for bound states of our EM
Hamiltonian. Thus, we will solve numerically the equation
(\ref{en}) for $E$, and the roots, if they exist, may lie in
principle inside the allowed and forbidden bands. However, the
numerical results could be well-handled if we have, a priori, some
analytical insight from the theory. To achieve this insight we
have to fall back upon the auxiliary constant-mass Schr\"odinger
equation of which we know very well the structure of the spectrum.
In the next section, we will obtain an image of our EM model in
this conventional Schr\"odinger language.

\section{Auxiliary constant-mass equation}\label{auxiliary}

Our purpose is to extend (\ref{lame}) onto the whole $z$-axis in
the form
\begin{equation}\label{se}
  -\ddot\chi(z)+\widetilde{V}(z)\chi (z)=E\chi (z),
\end{equation}
which can be viewed as a constant-mass Schr\"odinger equation for
a single particle of unit mass (according to our chosen scale
$\hbar^2=2$). Of course, equation (\ref{se}) plays the role of
auxiliary equation, and our aim is to extract general information
about its spectral properties, which could be used as a guide in
our numerical procedure for physical model (\ref{eigen1}). The
observation here is that this can be achieved by extending the
transformations (\ref{tri}) to the entire $z$-axis
\begin{equation}\label{tr}
  z=\int^x\sqrt{m(\tau)}d\tau, \qquad \chi(z)=\left [m(x(z))\right ]^{-1/4}\psi (x(z))\, .
\end{equation}
But there are two subtle points which should be taken into account
to obtain the correct expression for the Schr\"odinger potential
$\widetilde{V}(z)$. In the first place, since $m(x)$ is continuous,
the new coordinate $z(x)$ should also be continuous. It will then
follow from (\ref{tr}) that the constant-mass wave function $\chi
(z)$ is also continuous function of $z$. One can achieve this by
exploiting the arbitrariness in the indefinite integral in
(\ref{tr}). Indeed the explicit relation between $z$ and $x$ is
\begin{equation}\label{xz}
x(z) = \left\{
\begin{array}{ll}
 (z-\lambda_-)/\sqrt{m_0}, & -\infty<z<-K/2, \\ [1ex]
  \sn z/\cn z, & -K/2<z<K/2, \\ [1ex]
  (z-\lambda_+)/\sqrt{m_0}, & K/2<z<\infty,
  \end{array}
  \right.
\end{equation}
where $\lambda_{\pm}=\pm(K/2-\sqrt{m_0}x_0)$. The second point is
even more fundamental concerning the nature of $m'$ and $m''$ in
the whole region $x\in \mathbb{R}$. We have mentioned in
Subsec.~\ref{matching}  that due to the cancellation effect $m'$
will contain $\Theta$-discontinuity and consequently $m''$ will
produce $\delta$-discontinuity at the two
 junctions $x=\pm x_0$. Noting that $\delta(b x)=\delta(x)/|b|$,
  the Schr\"odinger potential $\widetilde{V}(z)$ may be expressed in the following form
\begin{equation}\label{sp}
  \hspace{-1.2cm}\widetilde{V}(z) = V_0 \, \widetilde{\Xi}(z)+\tilde{f}_{V}(z)[1-\widetilde{\Xi}(z)]
            +\left (\beta+\frac{1}{2}\right )(1+k')\left [\delta(z_+)+\delta(z_-)\right ],
\end{equation}
where
\begin{equation}\label{def6}
 \hspace{-1.2cm}\tilde{f}_{V}(z) =  2k^2 \left [ \sn^2z+\frac{\cn^2z}{\dn^2z} \right ]\, ; \qquad
 \widetilde{\Xi}(z) =  \Theta(-z_+)+\Theta(z_-), \quad z_{\pm}=z\pm K/2\, .
\end{equation}
The point to be noticed here is the $\beta$-dependence of the
shape of $\widetilde{V}(z)$ at the two junctions $z=\pm K/2$. It
should be kept in mind that although the auxiliary constant-mass
Schr\"odinger wave function $\chi(z)$ is continuous, its
derivative is not. In this sense the auxiliary Schr\"odinger
equation (\ref{se}) is not exactly in conventional form. One may
check readily from the matching condition (\ref{dc}) that
\begin{equation}
 \triangle \left [ \dot{\chi}(z) \right ]
   \Bigr|_{z=\pm K/2}= \left (\beta+\frac{1}{2}\right )(1+k')\chi^{\pm}.
\end{equation}
To understand the correct range of $E$ for bound states, we will
consider separately three cases: $\beta>-1/2$, $\beta=-1/2$ and
$\beta<-1/2$.

\begin{itemize}\label{range}

\item{\bf Case 1) $\mathbf{\boldsymbol{\beta>-1/2}}$}

In this case there will be a well, determined by one-period
associated Lam\'e potential $\tilde{f}_V(z)$ in the region
$(-K/2,K/2)$, bounded by two $\delta$-barriers, and outside this
region a constant potential $V_0$. Three possible situations may
arise (see Fig.~\ref{fig2}).
\begin{figure}[t]
\centering
\begin{pspicture}(-4.2,-0.8)(5,4.5)
 \psset{xunit=1.2cm} \psset{yunit=0.8cm}
\psline[linewidth=2pt]{->}(2,-0.5)(2,5.2) 
\psline[linewidth=2pt]{->}(-2,-0.5)(-2,5.2)
\pscurve{-}(1.99,2)(2,2)(4,2)
\pscurve{-}(-4,2)(-2,2)(-1.99,2) 
\pscurve{<->}(-4.5,-0.5)(0,-0.5)(4.5,-0.5)
\pscurve[linestyle=dashed,dash=3pt 2pt]{-}(-2,4.5)(-1.5,4.4)(-1,3.5)(0,3.2)(1,3.5)(1.5,4.4)(2,4.5)
\pscurve{-}(-2,2.5)(-1.5,2.4)(-0.8,0.5)(-0.5,-0.2)(0,-0.4)(0.5,-0.2)(0.8,0.5)(1.5,2.4)(2,2.5)
\pscurve[linestyle=dotted]{-}(-2,1.8)(-1.6,1.77)(-1,1.2)(0,0)(1,1.2)(1.6,1.77)(2,1.8)
\rput(-2,-0.9){\footnotesize{$-K/2$}}%
\rput(2,-0.9){\footnotesize{$K/2$}}%
\rput(4.5,-0.8){\footnotesize{$z$}}%
\rput(4,2.18){\footnotesize{$V_0$}}%
\rput(0,2){\footnotesize{$\tilde{f}_V(z)$}}%
\psline{->}(0,2.2)(0,3.2) \psline{->}(-0.4,2)(-1.2,2)
\psline{->}(0.1,1.8)(0.7,0.77)
\rput(0,5.3){{\footnotesize{$\delta$ barriers}}}%
\psline{->}(-0.73,5.15)(-1.95,4.8) \psline{->}(0.73,5.15)(1.95,4.8)
\end{pspicture}
\caption{Three possible positions of the constant-mass Schr\"odinger
potential $\widetilde{V}(z)$ are shown for $\beta>-1/2$. For
$\beta=-1/2$ $\delta$-barriers disappear, and
 for $\beta<-1/2$ there are two $\delta$-wells at the junctions.} \label{fig2}
\end{figure}
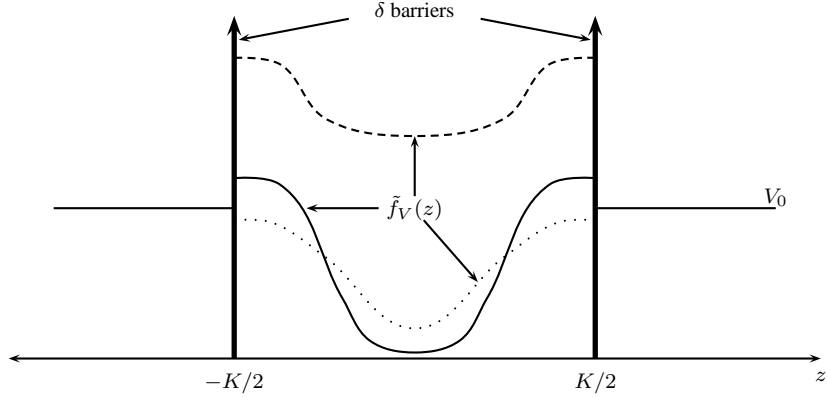
 Clearly there may be  bound states in the range $(\tilde{f}_V)_{min}<E<V_0$ if and only if
 $(\tilde{f}_V)_{min}<V_0$.

\item{\bf Case 2) $\mathbf{\boldsymbol{\beta=-1/2}}$}

In this case the $\delta$-barrier will disappear and for bound
states the conclusion is the same as before.

\item{\bf Case 3) $\mathbf{\boldsymbol{\beta<-1/2}}$}

In this situation the shape of the associated Lam\'e well provides
no restrictions from below on the bound state energies due to the
presence of two $\delta$-wells at the two junctions $z=\pm K/2$.
We thus conclude that the bound states may exist in the range
$-\infty<E<V_0$.

\end{itemize}

In the next section we will consider the limit $k\rightarrow 0$
and will write the explicit energy equation for this limit. Before
concluding this section it may be mentioned that in the
$k\rightarrow 0$ limit, $\tilde{f}_{V}(z)\equiv 0$ and
$\widetilde{V}(z)$ reduces to
\begin{equation}\label{sp0}
 \widetilde{V}(z)\rightarrow V_0 \, \widetilde{\Xi}(z)+\left (2\beta+1\right )\left [ \, \delta
(z_+)+\delta (z_-)\,\right ]\, ,
\end{equation}
where $z_{\pm}\rightarrow z\pm \pi/4$. Thus, for $\beta\geq -1/2$,
(Case 1 and 2 above)  bound states may exist in the range $0<
E<V_0$ and for $\beta< -1/2$ the possible range is
$-\infty<E<V_0$.

\section{A limiting case : $\mathbf{k\rightarrow 0}$}
It is well-known that the Jacobian elliptic functions degenerate
into trigonometric and hyperbolic functions in $k\rightarrow 0$
and $k\rightarrow 1$ limits. We have already pointed out that
$k\rightarrow 1$ limit is prohibited\footnote{But this limit
becomes interesting if different boundary conditions are adopted
\cite{agsolo}.} in our model, as this limit corresponds to the
nonphysical situation of vanishing mass for large $|x|$. In the
$k\rightarrow 0$ limit, the intermediate region shrinks up to the
interval $(-1,1)$ inside which the harmonic oscillator well $V(x)$
and the mass function $m(x)$ are
\begin{equation}\label{emp0}
\hspace{-1.5cm}\begin{array}{ll}
  V(x)=\left\{
  \begin{array}{ll}
    Cx^2+D ,& |x|\leq 1\\
    V_0 ,& |x|\geq 1
  \end{array}\right., &
  m(x)=\left\{
    \begin{array}{ll}
    (1+x^2)^{-2}, & |x|\leq 1\\
    1/4 ,& |x|\geq 1 \, .
  \end{array}\right.
\end{array}
\end{equation}

In the two infinite regions $-\infty<x<-1$ and $1<x<\infty$, the
solutions will be given by (\ref{tr1}) and (\ref{w1}), where
$\kappa,m_0$ and $V_0$ take the limiting forms
\begin{equation}\label{limdef}
 \hspace{-1.5cm}\kappa  \rightarrow  \left . \sqrt{V_0-E}\right /2,  \quad m_0\rightarrow 1/4\, ,
 \quad V_0 \rightarrow  5+8\left [\, \beta+2\alpha(\alpha+\beta+1)\right ].
\end{equation}
In the intermediate region, we have obtained the free-particle
equation
\begin{equation}
  -\ddot\chi (z)=E\chi (z), \qquad z\in(-\pi/4,\pi/4),
\end{equation}
where the transformations (\ref{tri}) now reduces to
\begin{equation}\label{tri0}
 x=\tan z, \qquad \chi (z)=(\sec z)\psi (x(z))\, .
\end{equation}
 Thus, the envelope wave function of (\ref{eigen1}) for $k\rightarrow 0$ acquires the form
\begin{equation}\label{w0}
  \psi (x) = \left \{
  \begin{array}{ll}
   \mathcal{N}^{-}e^{\kappa x}/2, & x<-1 \, , \\ [1ex]
   \displaystyle \frac{d_1e^{i\sqrt{E}\tan^{-1}(x)}+d_2e^{-i\sqrt{E}\tan^{-1}(x)}}{\sqrt{1+x^2}}, & |x|<1, \\ [2.5ex]
   \mathcal{N}^{+}e^{-\kappa x}/2, & x>1 \, .
  \end{array}
   \right.
\end{equation}

 As before, imposing the matching conditions, we obtain
homogeneous linear systems for $d_1,d_2$
\begin{eqnarray*}
  d_1e^{i\sqrt{E} \pi/4}[\mathcal{B}+i\sqrt{E}]+d_2e^{-i\sqrt{E} \pi/4}[\mathcal{B}-i\sqrt{E}]=0 \, ,
\\ [1ex]
  d_1e^{-i\sqrt{E}\pi/4}[\mathcal{B}-i\sqrt{E} ]+d_2e^{i\sqrt{E} \pi/4}[\mathcal{B}+i\sqrt{E}]=0\, .
\end{eqnarray*}
Thus the energy equation for bound states $(E<V_0)$ in the
$k\rightarrow 0$ limit is as follows
\begin{equation}\label{en0}
  (\mathcal{B}^2-E)\sin (\frac{\pi}{2}\sqrt{E})+2 \mathcal{B} \sqrt{E}  \cos (\frac{\pi}{2}\sqrt{E})=0 \, ,
\end{equation}
 where $\mathcal{B}$ is given by (\ref{def5}) for $k'\rightarrow
 1$. One may notice at once that $E=0$ is a trivial root of
 equation (\ref{en0}), but it must be rejected, because the
 solution $\psi(x)$ for the intermediate region $(-1,1)$, given in
 (\ref{w0}), is valid for $E\neq 0$. Indeed, for $E=0$ one may
 rewrite the solution as
\begin{equation}\label{w0d}
  \psi(x)=\frac{1}{\sqrt{1+x^2}}[d_1\tan^{-1}x+d_2], \quad -1<x<1,
\end{equation}
the expressions in two semi-infinite regions being the same as in
(\ref{w0}). In this case the matching conditions give the
following constraints for the existence of zero energy root
\begin{equation}\label{cond}
 \hspace{-1.9cm} V_0=(2\beta+1)^2, \: \beta<-\frac{1}{2}; \quad \mbox{or}\quad
   V_0=\left [2\beta+1+\frac{4}{\pi}\right ]^2, \: \beta<-\left (\frac{1}{2}+\frac{2}{\pi}\right ),
\end{equation}
and correspondingly in (\ref{w0d}) either $d_1=0$ or $d_2=0$. We
have checked that the constraints (\ref{cond}) are not satisfied
for different cases, so that zero energy state does not exist for
them.

\section{Numerical results for bound states}
 In this section we will solve numerically the energy equations (\ref{en}) [and (\ref{en0}) for $k\rightarrow 0$ limit]
 in the range $\widetilde{V}_{min}<E<V_0$, where $\widetilde{V}_{min}$ denotes the minimum of auxiliary
 potential $\widetilde{V}(z)$ given by (\ref{sp}) [and (\ref{sp0})]. It may be mentioned that the energy equations
 depend on two classes of parameters : i) ordering parameters $\alpha,\beta,\gamma$ connected by
 $\alpha+\beta+\gamma=-1$, and ii) elliptic modulus $k^2$ $(0\leq k^2<1)$ or the complementary modulus
 $k'^2=1-k^2$. Thus we have examined the roots of the energy equations for bound states as a function of
 both these two classes of parameters. Our strategy is to vary these parameters in such a way that it will
 cover some of the special forms of kinetic energy operator (\ref{kin}) mentioned in the introduction.

\subsection{One parameter family of kinetic energy operator for $\alpha=\gamma$}
 In this case kinetic energy operator (\ref{kin}) reduces to
 \begin{equation}\label{kin1}
 T_{EM}(x)=\frac{1}{2}\left ( m^{\alpha}pm^{\beta}pm^{\gamma} \right ), \quad 2\alpha+\beta=-1.
 \end{equation}
We will first obtain $E$ as a function of $k^2\in [0,1)$ for the two
values of $\beta$: $\beta=-1$ and $\beta=0$.

\vspace*{.5cm}
\noindent
\begin{minipage}[h]{1in}
 \textbf{Case $\mathbf{\boldsymbol{\beta=-1}}$}
 \end{minipage}\

\vspace*{.2cm}
 \noindent

For the choice $\beta=-1$, the kinetic energy operator (\ref{kin1})
takes the following form
\begin{equation}
 T_{EM}(x)=\frac{1}{2}\left ( p\frac{1}{m}p \right )\, .
\end{equation}
Although in this case  the upper and lower limits for bound states
are
\begin{equation}
 V_0=\frac{9k^2(1-k')}{4(1+k')}-(1+2k')+\frac{k^2}{4}\, ,
\end{equation}
and $\widetilde{V}_{min}=-\infty$, since $\beta<-1/2$ [see equation
(\ref{sp})], no bound states are obtained in this range.

\vspace*{.5cm}
 \noindent
 \begin{minipage}[h]{1in}
 \textbf{Case $\mathbf{\boldsymbol{\beta=0}}$}
\end{minipage}\

\vspace*{.2cm}
 \noindent

 This choice of $\beta$ corresponds to the kinetic energy operator
\begin{equation}
 T_{EM}(x)=\frac{1}{2}\left ( \frac{1}{\sqrt{m}}p^2\frac{1}{\sqrt{m}} \right )\, .
\end{equation}
Here, the range is
\begin{equation}
  V_0=\frac{5k^2(1-k')}{4(1+k')}+1+\frac{k^2}{4}, \qquad \widetilde{V}_{min}=2k^2 \, .
\end{equation}
In the interval $(\widetilde{V}_{min},V_0)$ there exist no bound
states for $k^2\in [0,1)$.

Next we will study the bound-state energies as a function of
$\beta\in [-2,2]$ for the two cases: $k^2=0$ and $k^2=0.5$.

\vspace*{.5cm}
 \noindent
 \begin{minipage}[h]{1in}
 \textbf{Case $\mathbf{k^2=0}$}
 \end{minipage}\

\vspace*{.2cm}
 \noindent

The upper limit $V_0$ has a parabolic dependence on $\beta$, given
by
\begin{equation}
  V_0=1-4\beta^2 \, ,
\end{equation}
and the lower limit is
\begin{equation}\label{vminc1}
  \widetilde{V}_{min} = \left \{
  \begin{array}{ll}
    0, & -\frac{1}{2}\leq\beta\leq 2,  \\ [1ex]
    -\infty, &  -2\leq\beta< -\frac{1}{2} \, .
  \end{array}
  \right.
\end{equation}
Our numerical calculation shows the existence of one bound state
near the top of the well for $-0.4\leq \beta \leq -0.1$ (see
Table~1).%
%
\begin{table}[t]\label{table1}
\center
 \caption[$E$ vs $\beta$ for $\alpha=\gamma,\, k^2=0$]
   {The bound state energies are calculated for $\beta\in [-2,2]$ and  $\alpha=\gamma,\, k^2=0$ along with
    $V_0,V_{min}$ ($x_0=1$). The bound states are observed only in the range $[-0.4,-0.1]$.}

 \vspace*{.2cm}
 \noindent
 \begin{tabular}{c@{\hspace{1cm}}c@{\hspace{1cm}}c@{\hspace{1cm}}c}\hline \hline
 $\beta$ & $V_{min}$ & $V_0$ & $E$  \\ \hline
 $-0.10$ & $0.80$ & $0.96$ & $0.94$ \\
 $-0.20$ & $0.60$ & $0.84$ & $0.81$ \\
 $-0.30$ & $0.40$ & $0.64$ & $0.62$ \\
 $-0.40$ & $0.20$ & $0.36$ & $0.35$ \\ \hline \hline
 \end{tabular}
 \end{table}

 \vspace*{.5cm}
 \noindent
 \begin{minipage}[h]{1in}
 \textbf{Case $\mathbf{k^2=0.5}$}
  \end{minipage}

\vspace*{.2cm}
 \noindent

 In this case also the upper limit $V_0$ has a
parabolic dependence on $\beta$, given by
\begin{equation}
  V_0=\frac{69-29\sqrt{2}}{20}-\frac{5}{4}\left ( \beta-\frac{2\sqrt{2}-1}{5}\right )^2\, ,
\end{equation}
while the lower limit is given by
\begin{equation}\label{vminc}
  \widetilde{V}_{min} = \left \{
  \begin{array}{ll}
    1, & -\frac{1}{2}\leq\beta\leq 2,  \\ [1ex]
    -\infty, & -2\leq\beta< -\frac{1}{2}\, .
  \end{array}
   \right.
\end{equation}
No bound states exist in this range.

\subsection{Two parameter family of kinetic energy operator for $\alpha\neq\gamma$}
\begin{figure}[ht]
 \rput(7.10,4.00){$\boldsymbol{\leftarrow}$\footnotesize{$V_0$}}%
\rput(0.60,6.20){$\mathbf{E}$}%
\rput(7.8,1.90){$\boldsymbol{\uparrow}$}%
\rput(8.0,1.5){\footnotesize{$\widetilde{V}_{min}$}}%
\rput(9.9,0.60){$\mathbf{k^2}$}%
\centering \epsfig{file=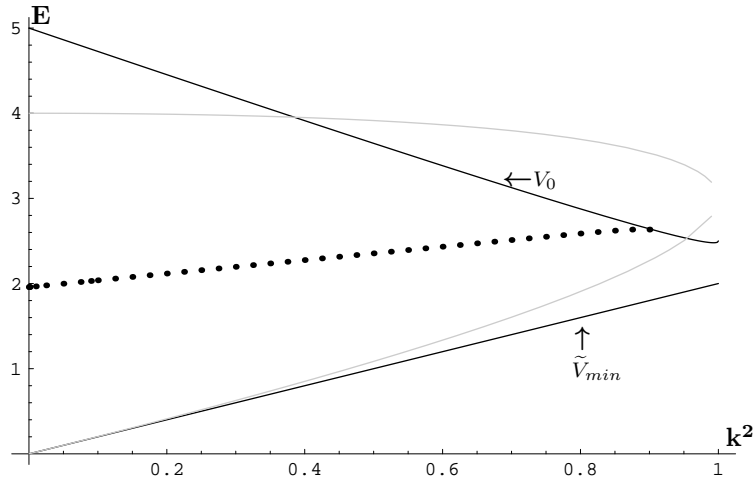, width=10cm} \caption{\small
Bound state energies
 as a function of $k^2$ are shown by dots for the case $\alpha=-1$, $\beta=\gamma=0$. The lower and upper gray
 curves represent first two band-edge energies $E^{(j)},j=0,1$ for the corresponding constant-mass periodic
 associated Lam\'e potential.}   \label{fig3}
\end{figure}

We will find $E$ as a function of $k^2$ for $\alpha=-1,\beta=0$
which yields following kinetic energy operator
\begin{equation}
    T_{EM}(x)=\frac{1}{4}\left [\frac{1}{m}p^2+p^2\frac{1}{m}\right ].
\end{equation}
The bounds are given by
\begin{equation}
  V_0=\frac{3k^2(1-2k')}{2(1+k')}+4k'+1, \qquad \widetilde{V}_{min}=2k^2\, .
\end{equation}
 We have obtained one bound state of the potential (see Table~2) for $0\leq k^2\leq 0.90$.%
%
 \begin{table}[t]
 \center
 \caption[$E$ vs $k^2$ for $\alpha=-1,\beta=0$]
   {Bound state energies are provided for $k^2\in[0,1)$ where $\alpha=-1,\beta=\gamma=0$. The point $x_0$ and
   $V_0,V_{min}$ for $V (x)$ are shown in each case. Bound states start to appear from $k^2\leq 0.90$.}

 \vspace*{.2cm}
 \noindent
 \begin{tabular}{c@{\hspace{1cm}}c@{\hspace{1cm}}c@{\hspace{1cm}}c@{\hspace{1cm}}c} \hline \hline
 $k^2$ & $x_0$ & $V_{min}$ & $V_0$ & $E$  \\ \hline
 $0.90$ & $1.78$ & $0.32$ & $2.64$ & $2.63$ \\
 $0.80$ & $1.50$ & $0.40$ & $2.88$ & $2.59$ \\
 $0.70$ & $1.35$ & $0.48$ & $3.13$ & $2.51$ \\
 $0.60$ & $1.26$ & $0.55$ & $3.38$ & $2.43$ \\
 $0.50$ & $1.19$ & $0.62$ & $3.65$ & $2.35$ \\
 $0.40$ & $1.14$ & $0.70$ & $3.91$ & $2.28$ \\
 $0.30$ & $1.09$ & $0.78$ & $4.18$ & $2.20$\\
 $0.20$ & $1.06$ & $0.85$ & $4.45$ & $2.12$ \\
 $0.10$ & $1.03$ & $0.92$ & $4.73$ & $2.04$ \\
 $0.00$ & $1.00$ & $1.00$ & $5.00$ & $1.96$ \\ \hline \hline
 \end{tabular}
 \end{table}
 The effective mass potential $V(x)$ maintains the shape of the
well for all $k$, and in particular
  for $k=0$ it becomes well-known harmonic oscillator inside the interval $(-x_0,x_0)$, the minima being at $x=1$.
 It is interesting
 to observe that the bound state exists inside
the first allowed band of periodic constant-mass associated Lam\'e
potential (see Fig.~\ref{fig3}).

The dependence of bound-state energy $E$ on $\beta$ will now be
calculated again for two cases: $k^2=0$ and $k^2=0.35$, while
$\beta$ will vary in the same interval $[-2,2]$ and $\alpha$ will
be fixed as $\alpha=-1$.

 \vspace*{.5cm}
 \noindent
 \begin{minipage}[h]{2in}
 \textbf{Case $\mathbf{k^2=0,\boldsymbol{\alpha=-1,\gamma=-\beta}}$}
 \end{minipage}

\vspace*{.2cm}
 \noindent

\begin{figure}[ht]
\rput(4.80,4.50){$\boldsymbol{\leftarrow}$\footnotesize{$V_0$}}%
\rput(6.70,6.00){$\mathbf{E}$}%
\rput(8.8,0.65){$\boldsymbol{\downarrow}$}%
\rput(8.8,1.00){\footnotesize{$\widetilde{V}_{min}$}}%
\rput(9.6,0.60){$\boldsymbol{\beta}$}%
 \centering \epsfig{file=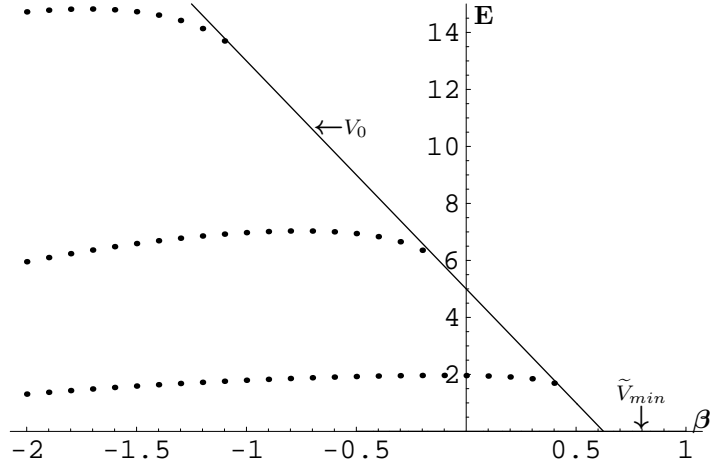, width=10cm} \caption{\small Bound state energies (dots) as
 a function of $\beta$ for
$\alpha\neq\gamma$, $\alpha=-1$, $k^2=0$. } \label{fig4}
\end{figure}

Here the upper limit $V_0$ is linearly dependent on $\beta$ and is
given by
\begin{equation}
  V_0=5-8\beta,
\end{equation}
and the lower limit $\widetilde{V}_{min}$ is same as given by
(\ref{vminc1}). The results, shown in Fig.~\ref{fig4}, are the
following
\begin{itemize}
  \item No bound states for $0.5\leq\beta\leq 2$,
  \item One bound state for $-0.1\leq\beta\leq 0.4$,
  \item Two bound states for $-1.0\leq \beta\leq -0.2$,
  \item Three bound states for $-2.0\leq \beta\leq -1.1$.
\end{itemize}

One can notice that for $\beta=0.4$, $V(x)\equiv 1.8$ is a
constant potential (see Table~3), but still there exists one
bound state.%
%
 \begin{table}[t]
 \center
 \caption[$E$ vs $\beta$ for $\alpha=-1\neq \gamma,k^2=0.35$]
   {The bound state energies as a function of $\beta\in [-2,2]$ for $\alpha=-1\neq\gamma, k^2=0$ are given with
    $V_{min}$ and $V_0$ ($x_0=1$). The asterisk in $\beta$ indicates that $V (x)$ is constant potential.}

 \vspace*{.2cm}
 \noindent
 \begin{tabular}{c@{\hspace{1cm}}c@{\hspace{1cm}}c@{\hspace{1cm}}c}\hline \hline
 $\beta$ & $V_{min}$ & $V_0$ & $E$  \\ \hline
 \hspace{2pt}$0.40^*$ & $1.80$ & $1.80$ & $1.69$ \\
 \hspace{2pt}$0.20$ & $1.40$ & $3.40$ & $1.91$ \\
 \hspace{2pt}$0.00$ & $1.00$ & $5.00$ & $1.96$ \\
 $-0.20$ & $0.60$ & $6.60$ & $1.96; 6.36$ \\
 $-0.40$ & $-0.20$ & $8.20$ & $1.94; 6.83$ \\
 $-0.60$ & $-0.20$ & $9.80$ & $1.91; 7.00$ \\
 $-0.80$ & $-0.60$ & $11.40$ & $1.86; 7.03$ \\
 $-1.00$ & $-1.00$ & $13.00$ & $1.79; 6.97$ \\
 $-1.20$ & $-1.40$ & $14.60$ & $1.72; 6.86; 14.14$ \\
 $-1.40$ & $-1.80$ & $16.20$ & $1.64; 6.70; 14.61$ \\
 $-1.60$ & $-2.20$ & $17.80$ & $1.54; 6.48; 14.79$ \\
 $-1.80$ & $-2.60$ & $19.40$ & $1.43; 6.24; 14.81$ \\
 $-2.00$ & $-3.00$ & $21.00$ & $1.31; 5.95; 14.72$ \\ \hline \hline
 \end{tabular}
 \end{table}
However, we have already explained in Sec.~\ref{auxiliary}  that
this apparently strange behavior is perfectly consistent with
constant-mass Schr\"odinger picture. In this context it may be
mentioned that the existence of bound states for constant
potential was also noticed in Ref.~\cite{bag1} for
$m(x)=\mbox{sech}^2qx$.

\vspace*{.5cm}
 \noindent
 \begin{minipage}[h]{3in}
 \textbf{Case $\mathbf{k^2=0.35,\boldsymbol{\alpha=-1,\gamma=-\beta}}$}
 \end{minipage}

\vspace*{.2cm}
 \noindent

 This case also corresponds to linear dependence of $V_0$ on $\beta$
\begin{equation}
  V_0=4.05-5.16 \: \beta \, ,
\end{equation}
and the lower limit is given by
\begin{equation}
\widetilde{V}_{min} = \left \{
  \begin{array}{ll}
   0.7, & \beta\geq -1/2, \\ [1ex]
  -\infty, & \beta< -1/2 \, .
  \end{array}
   \right.
\end{equation}
The results shown in Table~4  are :
%
 \begin{table}[t]
 \center
 \caption[$E$ vs $\beta$ for $\alpha=-1\neq \gamma,k^2=0.35$]
   {The bound state energies as a function of $\beta\in [-2,2]$ for $\alpha=-1\neq\gamma, k^2=0.35$ are given with
    $V_{min}$ and $V_0$ ($x_0=1.11$).}

 \vspace*{.2cm}
 \noindent
 \begin{tabular}{c@{\hspace{1cm}}c@{\hspace{1cm}}c@{\hspace{1cm}}c}\hline \hline
 $\beta$ & $V_{min}$ & $V_0$ &  $E$  \\ \hline
 \hspace{2pt}0.40 & 1.96 & 1.98 & $1.98$ \\
 \hspace{2pt}0.20 & 1.35 & 3.01 & $2.20$ \\
 \hspace{2pt}0.00 & 0.74 & 4.05 & $2.24$ \\
 -0.20 & 0.13 & 5.08 & $2.23$ \\
 -0.40 & -0.48 & 6.11 & $2.20; 5.67$ \\
 -0.60 & -1.09 & 7.14 & $2.16; 5.86$ \\
 -0.80 & -1.70 & 8.18 & $2.10; 5.89$ \\
 -1.00 & -2.31 & 9.21 & $2.04; 5.83$ \\
 -1.20 & -2.92 & 10.24 & $1.96; 5.71$ \\
 -1.40 & -3.53 & 11.27 & $1.87; 5.54; 11.04$ \\
 -1.60 & -4.14 & 12.31 & $1.77; 5.34; 11.35$ \\
 -1.80 & -4.75 & 13.34 & $1.66; 4.96; 11.43$ \\
 -2.00 & -5.36 & 14.37 & $1.53; 4.82; 11.40$ \\\hline \hline
 \end{tabular}
 \end{table}
\begin{itemize}
  \item No bound states for $0.5\leq\beta\leq 2$,
  \item One bound state for $-0.2\leq\beta\leq 0.4$,
  \item Two bound states for $-1.2\leq \beta\leq -0.3$,
  \item Three bound states for $-2\leq \beta\leq -1.3$.
\end{itemize}

\begin{figure}[ht]
\rput(4.30,4.50){$\boldsymbol{\leftarrow}$\footnotesize{$V_0$}}%
\rput(6.70,6.00){$\mathbf{E}$}%
\rput(8.8,0.95){$\boldsymbol{\downarrow}$}%
\rput(8.8,1.30){\footnotesize{$\widetilde{V}_{min}$}}%
\rput(9.6,0.60){$\boldsymbol{\beta}$}%
\centering \epsfig{file=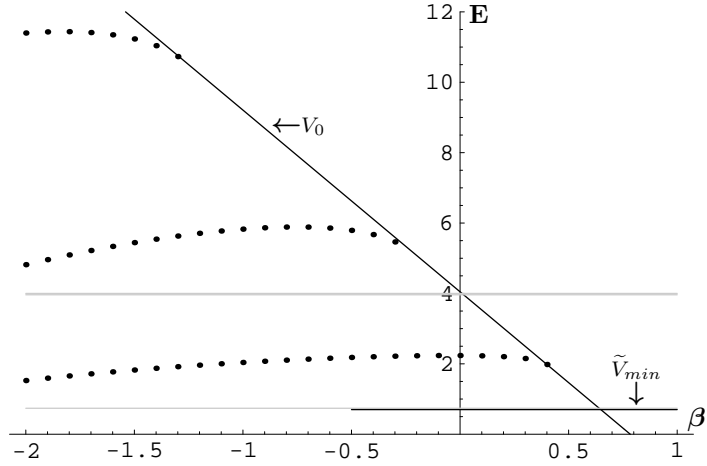, width=10cm} \caption{\small
Bound state energies (dots) as a function of $\beta$ are plotted
for $\alpha\neq\gamma$, $\alpha=-1$, $k^2=0.35$. The gray lines
represent band-edge energies of corresponding associated Lam\'e
potential.} \label{fig5}
\end{figure}

 Once again we have observed that the ground state lies
inside the allowed band and higher excited states are inside the
continuum (see Fig.~\ref{fig5}) for
the constant-mass periodic associated Lam\'e potential
characterized by $\mu=\nu=1;\, k^2=0.35$.

\section{Some features of bound-state wave functions}
In this section we will present some interesting properties of the
wave functions $\psi(x)$ that can be derived from the numerical
results of the previous section with very good accuracy. A part of
these properties, which seems to be difficult to prove
analytically, does not concern the specific physical problem. We
recall that $\psi(x)$ is given by (up to the normalization factor)
\begin{equation}
 \psi(x) = \left \{
  \begin{array}{ll}
   m_0^{1/4}\chi^-e^{\kappa (x_0+x)}, & -\infty<x<-x_0 \, , \\ [1ex]
   f^{1/4}_m(x)\chi(z(x)), & -x_0<x<x_0 \, , \\ [1ex]
   m_0^{1/4}\chi^+e^{\kappa (x_0-x)}, & x_0<x<\infty \, ,
  \end{array}
   \right.
 \end{equation}
 where $\chi(z(x))=\chi_2(z(x))+(d_1/d_2)\chi_1(z(x))$ and
 $\chi^{\pm}\equiv\chi(\pm K/2)$. Note that for $k\neq 0$,
 $\chi_i(z(x))$ are given by (\ref{ls}) for $x(z)=\sn \, z/\cn \, z$, and for
 $k=0$, $\chi_{1,2}(z(x))=\exp [\pm i\sqrt{E}z]$ for $x(z)=\tan
 z$. The quantity $\kappa$ is to be computed from (\ref{def3}) for
 $E=E_n$, $E_n$ being the bound-state energies of our EM Hamiltonian.
 In the following we will use the notation
\begin{equation}
  \kappa_n=\sqrt{m_0(V_0-E_n)}, \: n=0,1,2 \, .
\end{equation}

\subsection{Ground state}
 The two quantities $a_1$ and $a_2$ in (\ref{ls}) are complex and related by
\begin{equation}
  a_1=-a_2^* \, .
\end{equation}
The real and imaginary part of $\chi_i(z)$ are respectively odd
and even functions. Moreover $\chi_i(z)$ satisfy following
interesting relations with very high accuracy
\begin{equation}\label{def19}
  \chi_1(z)=\frac{d_2}{d_1}\chi_2(-z), \quad \chi_i(z)=-\chi_i^*(-z) \, ,
\end{equation}
where the ratio $d_2/d_1$ is a positive number. It will then
follow that $\chi(z)$ is an even function and
\begin{equation}
  \chi(z)=\chi(-z)\propto \im[\chi_1(z)]\equiv v(z)\, .
\end{equation}
The ground state wave function is
\begin{equation}
 \psi_0(x) = \left \{
  \begin{array}{ll}
   \mathcal{C}_0 m_0^{1/4}
             v\left (\frac{K}{2}\right )e^{\kappa_0(x_0+x)}, & x<-x_0 \, , \\ [1ex]
   \mathcal{C}_0 f^{1/4}_m(x) v(z(x)), & |x|<x_0 \, , \\ [1ex]
   \mathcal{C}_0 m_0^{1/4}
    v\left (\frac{K}{2}\right )e^{\kappa_0(x_0-x)}, &  x>x_0 \, .
   \end{array}
    \right.
 \end{equation}
It is straightforward to compute the normalization constant
$\mathcal{C}_0$ from (\ref{nc})
\begin{equation}
 \hspace{-.1cm}\frac{1}{\mathcal{C}_0}=\left [
  \frac{v^2(K/2)}{\sqrt{V_0-E_0}}+\int^{x_0}_{-x_0} \sqrt{f_m(x)}v^2(z(x))dx \right ]^{1/2}.
\end{equation}

\subsection{First excited state}
 The quantity $a_1$ is
purely imaginary while the other quantity $a_2$ is complex and
they are related by
\begin{equation}
  \im[a_1]=\im[a_2]\, .
\end{equation}
The relation between $\chi_1(z)$ and $\chi_2(z)$ are
\begin{equation}
  \chi_1(z)=-\frac{d_2}{d_1}\, \chi_2(-z)\, ,
\end{equation}
where $d_2/d_1$ is a positive number. Clearly this implies that
\begin{equation}
  \chi(-z)=-\chi(z)=\mbox{ complex }.
\end{equation}
We have checked numerically with the same high accuracy that
\begin{equation}
  \re[\chi(z)]\propto \im[\chi(z)]\, ,
\end{equation}
and of course no degeneracy exists for $E=E_1$. The wave function
is
\begin{equation}
 \psi_1(x) = \left \{
  \begin{array}{ll}
    -\mathcal{C}_1 m_0^{1/4}
              \re[\chi^+]e^{\kappa_1(x_0+x)}, & x<-x_0 \, , \\[1ex]
  \mathcal{C}_1 f^{1/4}_m(x) \re[\chi(z(x))], & |x|<x_0 \, , \\ [1ex]
  \mathcal{C}_1 m_0^{1/4}
         \re[\chi^+]e^{\kappa_1(x_0-x)}, & x>x_0 \, .
  \end{array}
   \right.
 \end{equation}

\subsection{Second excited state}
 The two quantities $a_1$
and $a_2$ have same properties as in the above case. The relation
between $\chi_1(z)$ and $\chi_2(z)$ reads in this case
\begin{equation}
  \chi_1(z)=\frac{d_2}{d_1}\chi_2(-z), \quad \left (\, \frac{d_2}{d_1}<0 \right )\, .
\end{equation}
Thus, here also $\chi(z)$ is complex and
\begin{equation}
  \chi(-z)=\chi(z)\, ,
\end{equation}
but
\begin{equation}
  \re[\chi(z)]\propto \im[\chi(z)]\, .
\end{equation}
The wave function is
\begin{equation}
 \psi_2(x) = \left \{
  \begin{array}{ll}
   \mathcal{C}_2 m_0^{1/4}
             \re[\chi^+]e^{\kappa_2(x_0+x)}, & x<-x_0 \, , \\[1ex]
  \mathcal{C}_2 f^{1/4}_m(x) Re[\chi(z(x))], & |x|<x_0 \, , \\ [1ex]
  \mathcal{C}_2 m_0^{1/4}
   Re[\chi^+]e^{\kappa_2(x_0-x)}, & x>x_0 \, .
  \end{array}
   \right.
 \end{equation}

The normalization constants $\mathcal{C}_n,n=1,2$ may be expressed
as
\begin{equation}
  \frac{1}{\mathcal{C}_n}=\left [\frac{(\re[\chi^+])^2}{\sqrt{V_0-E_n}}+
       \int^{x_0}_{-x_0} \sqrt{f_m(x)}(\re[\chi(z(x))])^2 dx\right ]^{1/2}.
\end{equation}

\subsection{The  limiting case for $k\rightarrow 0$}

In the limiting case for $k\rightarrow 0$ the relation between $x$
and $z$ is very simple [see equation (\ref{tri0})], and it is
convenient to express the corresponding properties in the variable
$x$. Here we will denote $\chi_i(z(x))$ and $\chi(z(x))$ by
$\chi_i(x)$ and $\chi(x)$. The relation between $\chi_1(x)$ and
$\chi_2(x)$ is
  \begin{equation}
  \chi_1(-x)=\chi_2(x) \, .
  \end{equation}
  Let us introduce the parameters
\begin{equation}
\theta_n=\frac{\pi}{4}\sqrt{E_n}\, , \qquad n=0,1,2 \, .
\end{equation}

  \vspace*{.5cm}
  \noindent
 \begin{minipage}[h]{4in}
 Ground and second excited states ($k=0$):
 \end{minipage}

\vspace*{.2cm}
 \noindent

  In both cases the ratio $d_2/d_1$ is unity. It then follows
from the definition of $\chi(x)$ that it is even function given by
\begin{equation}
  \chi(-x)=\chi(x)=2\cos (\sqrt{E}\tan^{-1}x)\, .
\end{equation}
The wave functions are
\begin{equation}
 \psi_n(x) = \left \{
  \begin{array}{ll}
   \mathcal{C}_n\sqrt{2}\cos \theta_n e^{\kappa_n(1+x)}, & x<-1 \\ [1ex]
   \frac{2\mathcal{C}_n}{\sqrt{1+x^2}}\cos (\sqrt{E_n}\tan^{-1}x), & |x|<1 \\[1ex]
    \mathcal{C}_n\sqrt{2}\cos \theta_n e^{\kappa_n(1-x)}, & x<-1 \, ,
   \end{array}
    \right.
\end{equation}
for $n=0,2$.

 \vspace*{.5cm}
 \noindent
 \begin{minipage}[h]{4in}
 First excited state ($k=0$):
 \end{minipage}

\vspace*{.2cm}
 \noindent

  The ratio $d_2/d_1$ is equal to $-1$
so that $\chi(x)$ is an odd function given by
\begin{equation}
  \chi(x)=-\chi(-x)=-2i\sin (\sqrt{E}\tan^{-1}x) \, .
\end{equation}
The wave function is
\begin{equation}
 \psi_1(x) = \left \{
  \begin{array}{ll}
    -\mathcal{C}_1\sqrt{2}\sin \theta_1 e^{\kappa_1(1+x)}, & x<-1 \\[1ex]
   \frac{2\mathcal{C}_1}{\sqrt{1+x^2}}\sin (\sqrt{E_1}\tan^{-1}x), & |x|<1 \\[1ex]
    \mathcal{C}_1\sqrt{2}\sin \theta_1 e^{\kappa_1(1-x)}, & x<-1 \, ,
   \end{array}
    \right.
\end{equation}

 In all
the three cases the normalization constants $\mathcal{C}_n$ may be
expressed as
\begin{equation}
  \frac{1}{\mathcal{C}_n}=\left [\pi+2\left (\frac{1+(-1)^n\cos
  2\theta_n}{\sqrt{V_0-E_n}}+(-1)^n\frac{\sin
  2\theta_n}{\sqrt{E_n}}\right )\right ]^{1/2}.
\end{equation}

 \section{Conclusion}
 In this article we have proposed a new solvable model wherein the potential and effective mass are
 rational functions of spatial coordinate. The
 novel feature of our model is that it may be mapped to well-known periodic associated Lam\'e
 potential in constant-mass scenario. This fact is clearly responsible for solvability of
 the model. It may be mentioned that the list of solvable models in
 effective mass framework is rather short. Our work has
 definitely enhanced this set by introducing a wide class of rational
 potential and mass functions. The important difference of
 our work compared to recent efforts is that we consider the
 variation of mass inside a finite region, and both potential and
 mass are constant outside. The advantage of considering variation
 of the mass inside a finite interval instead of full line is that
 the mass remains finite and non-zero everywhere, as it should be.

 We have examined the bound-state spectrum for different
 values of the ordering parameters $\alpha,\beta,\gamma$ and elliptic modulus $k^2$, and
 have discussed the properties of the corresponding wave functions. It is observed for both
 $\alpha=\gamma$ and $\alpha\neq\gamma$ cases that
 the ordering parameter $\beta$ has a critical value above which no bound state exists. But number
 of levels increases with decreasing values of $\beta$. Some peculiarities have been
 noted in the spectral properties [e.g. the existence of bound states for constant potential
 $V(x)$] in contrast to the conventional situation where mass is
 constant. The qualitative observation is that the discrete energy levels for
 bound states of our EM Hamiltonian for nonzero $k$ lie only inside the allowed band and in the
 continuum for corresponding constant-mass periodic associated
 Lam\'e Hamiltonian. Moreover, as a by-product of our numerical procedure we have
 found some new curious relations for the solutions $\chi_i(z)$ of associated Lam\'e equation,
 which seem to be universal. For example the
 form of the second relation of (\ref{def19}) for the solutions $\chi_i(z)$
 at $E=E_0$ (inside the allowed band of corresponding associated Lam\'e potential)
  is not related with our physical problem.

  Some direct generalizations of our model are possible. For instance, the application
  of SUSY transformations to enlarge the class of rational models and the inclusion of
  scattering states with $E\geq V_0$ are under consideration. Further
  it will be interesting
  to investigate the validity of the above mentioned relations between the solutions of associated
  Lam\'e equation for other values of
  $E$. The special exploration of $k\rightarrow 1$ limit of our
  model also seems to be promising \cite{agsolo}.

 \section*{Acknowledgements}
 A.G. acknowledges Spanish Ministry of Foreign Affairs for his research
 grant 0000147287 and the authorities of City College, Calcutta for study leave.
 M.V.I. is supported by the sabbatical grant SAB2004-0143 of the Spanish Ministry
 of Education and by Russian grants RFBR 06-01-00186-a, RNP 2.1.1.1112 and RSS-5538.2006.2.
 The work of L.M.N. is supported by the
 Spanish MEC (MTM2005-09183) and Junta de Castilla y Le\'on (Excellence project VA013C05).

 \appendix
 \section{}
\label{appendix}

 In the following we will mention basic definitions and the relations involving
  elliptic functions, which we use in the text (for more details, see [59--61]. 
  Consider two real numbers $k^2$ and $k'^2$ such that
\begin{equation}
  k^2\in (0,1), \qquad k'^2=1-k^2.
\end{equation}
These two numbers are called respectively elliptic modulus and
complementary modulus, and are basic parameters in the constructions
of elliptic functions. The amplitude function is defined by
\begin{equation}
  \varphi (z,k)=\am(z,k), \qquad z(\varphi,k)=\int_0^\varphi\frac{d\tau}{\sqrt{1-k^2\sin^2\tau}}\, .
\end{equation}
From here the three Jacobian elliptic functions are defined by
\begin{equation}
  \sn (z,k)=\sin \varphi\, , \quad  \cn(z,k)=\cos\varphi\, , \quad \dn(z,k)=d\varphi/dz \, .
\end{equation}
These functions are called sine-amplitude, cosine-amplitude and
delta-amplitude respectively. For simplicity, in equation
(\ref{lame}) in the text, and also in the following, we have
suppressed the explicit modular dependence and write simply $\sn \, z,
\cn z, \dn z$. These are doubly-periodic functions of periods
$(4K,2iK')$, $(4K,4iK')$ and $(2K,4iK')$ respectively, and are
usually defined for a complex variable $z$. Nevertheless in our case
$z$ is always real. The quarter-periods $K$ and $K'$ are the real
numbers given by
\begin{equation}
  K(k)\equiv K=z(\pi/2,k), \quad K'(k)\equiv K'=K(k')\, .
\end{equation}
$K$ is called complete elliptic integral of second kind. Noting
the following relations
\begin{eqnarray}
 \sn (z+K)=\frac{\cn \, z}{\dn \, z}\, , &  \quad \cn(z+K)= -k'\frac{\sn \, z}{\dn \, z}\, , \quad
      \dn(z+K)= \frac{k'}{\dn \, z}\, , \\[1ex]
 \sn (z+2K)=-\sn \, z \, , &  \quad \cn(z+2K)=-\cn \, z \, , \quad \: \: \: \dn(z+2K)=\dn \, z\, ,
\end{eqnarray}
we see that the associated Lam\'e potential in equation
(\ref{lame}) is $2K$-periodic or $K$-periodic, according as
$\mu\neq\nu$ or $\mu=\nu$. The function  $\sn z$ is odd with a
simple zero at $z=0$, while $\cn z,\dn z$ are even functions; $\cn
z$ has a simple zero at $z=K$, but $\dn z$ has no zeros for real
 $z$. The precise value of the point $x_0$ is obtained as
 $x_0=1/\sqrt{k'}$ from the values
\begin{equation}
  \sn  \left (\frac{K}{2} \right )=\frac{1}{\sqrt{1+k'}}\, , \qquad
  \cn \left (\frac{K}{2} \right )=\frac{\sqrt{k'}}{\sqrt{1+k'}}\, .
\end{equation}
Some other relevant relations are
\begin{equation}
  \sn^2z+\cn^2z=1 \, , \qquad \dn^2z+k^2\sn^2z=1 \, ,
\end{equation}
\begin{equation}
 \sn'z=\cn \, z \,\dn \, z\, , \quad \cn'z=-\sn \, z\,\dn \, z\, , \quad \dn'z=-k^2\sn \, z \,\cn \, z\, .
\end{equation}
In the two limits $k\rightarrow 1$ and $k\rightarrow 0$, the
elliptic functions degenerate into hyperbolic and trigonometric
functions
\begin{eqnarray}
  \sn z \renewcommand{\arraystretch}{.5}
           \hspace{-2pt}  \begin{array}{l}
  \scriptstyle{k \! \rightarrow \! 1} \\ \longrightarrow \\ \scriptstyle{k \! \rightarrow \! 0}
              \end{array}
                \hspace{-2pt} \left \{ \begin{array}{l}
                   \hspace{-2pt} \tanh z \\ \hspace{1pt} \\ \hspace{-2pt}\sin z
                          \end{array} \right . \hspace{-2pt}, &
             \quad   \cn z \renewcommand{\arraystretch}{.5}
            \hspace{-2pt} \begin{array}{l}
  \scriptstyle{k \! \rightarrow \! 1} \\ \longrightarrow \\ \scriptstyle{k \! \rightarrow \! 0}
             \end{array}
           \hspace{-2pt}\left \{ \begin{array}{l}
                    \hspace{-2pt}\mbox{sech}z \\ \hspace{1pt}\\ \hspace{-2pt}\cos z
                          \end{array} \right . \hspace{-2pt}, &
               \quad    \dn z \renewcommand{\arraystretch}{.5}
            \hspace{-2pt} \begin{array}{l}
  \scriptstyle{k \! \rightarrow \! 1} \\ \longrightarrow \\ \scriptstyle{k \! \rightarrow \! 0}
             \end{array}
                   \hspace{-2pt}\left \{ \begin{array}{l}
                      \hspace{-2pt}\mbox{sech}z \\ \hspace{1pt} \\ \hspace{-2pt}1
                           \end{array} \right .\hspace{-5pt} ,
  \end{eqnarray}
 \renewcommand{\arraystretch}{1}
 where the quarter-periods go over
 \begin{eqnarray}
  K \renewcommand{\arraystretch}{.5}
           \hspace{-2pt}  \begin{array}{l}
  \scriptstyle{k \! \rightarrow \! 1} \\ \longrightarrow \\ \scriptstyle{k \! \rightarrow \! 0}
              \end{array}
                \hspace{-2pt} \left \{ \begin{array}{l}
                   \hspace{-2pt} \infty \\ \hspace{1pt} \\ \hspace{-2pt}\pi/2
                          \end{array} \right . \hspace{-2pt}, &
                \qquad K' \renewcommand{\arraystretch}{.5}
            \hspace{-2pt} \begin{array}{l}
  \scriptstyle{k \! \rightarrow \! 1} \\ \longrightarrow \\ \scriptstyle{k \! \rightarrow \! 0}
             \end{array}
           \hspace{-2pt}\left \{ \begin{array}{l}
                    \hspace{-2pt}\pi/2 \\ \hspace{1pt}\\ \hspace{-2pt}\infty
                          \end{array} \right . \hspace{-2pt}. &
  \end{eqnarray}
  \renewcommand{\arraystretch}{1}

  Weierstrass elliptic function $\wp (z;g_2,g_3)\equiv \wp (z)$
  is defined by
  \begin{equation}
  \wp (z)=\frac{1}{z^2}+\sum_{m,n}\hspace{-1pt}^{'}
  \left [ \frac{1}{(z-2m\omega_1-2n\omega_3)^2}-\frac{1}{(2m\omega_1+2n\omega_3)^2} \right ],
  \end{equation}
  where the symbol $\sum'$ means summation over all
  integral values of $m,n$ except $m=n=0$, $\omega_1$ and $\omega_3$
  being half-periods of $\wp (z)$. In our case, these are defined
  through
\begin{equation}\label{om}
  \omega_1=K\, , \qquad \omega_3=i K'\, .
\end{equation}
and the invariants $g_2,g_3$ are given by
 \begin{equation}
 g_2=\frac{4}{3}(k^4-k^2+1)\, , \quad
 g_3=\frac{4}{27}(k^2-2)(2k^2-1)(k^2+1)\, .
 \end{equation}
 The above choice corresponds to the case when the discriminant
 $\Delta=g_2^3-27g_3^2>0$, and so the three numbers
 $\wp(\omega_i)\equiv e_i, \, i=1,2,3$ are always real ($e_1>e_2>e_3$),
 where $\omega_2=\omega_1+\omega_3$. It may be mentioned that
 $e_i$ are the three roots of the equation
 \begin{equation}
 4t^3-g_2t-g_3=0\, ,
 \end{equation}
 and, for simplicity, we have chosen the scale $e_1-e_3=1$.
 Weierstrass elliptic function is an even function; its derivative
 $\dot{\wp}(z)$ is an odd elliptic function with the same periods
 and satisfy following identity
 \begin{equation}
 \dot{\wp}^2(z)=4\:\prod^3_{i=1}\left [\,\wp (z)-e_i \,\right ].
 \end{equation}

 The relations between Weierstrass elliptic function and Jacobian
 elliptic functions, with our choice, are
 \begin{equation}
  \wp(z)=e_1+\frac{\cn^2z}{\sn^2z}=e_2+\frac{\dn^2z}{\sn^2z}=e_3+\frac{1}{\sn ^2z}\, .
 \end{equation}
 Weierstrass zeta function $\zeta (z;g_2,g_3)\equiv \zeta (z)$ and
 sigma function $\sigma (z;g_2,g_3)\equiv \sigma (z)$ are
 quasi-periodic functions defined by
\begin{equation}\label{sig}
  \dot{\zeta}(z)=-\wp (z)\, , \qquad \frac{\dot{\sigma} (z)}{\sigma (z)}=\zeta (z)\, ,
\end{equation}
and satisfy the following relations
\begin{equation}\label{odd}
 \begin{array}{ll}
  \zeta (-z)=-\zeta (z)\, , &
  \zeta(z+2\omega_i)=\zeta(z)+2\zeta(\omega_i)\, ,\\[1ex]
  \sigma (-z)=-\sigma (z)\, , & \sigma(z+2\omega_i)=-\sigma(z)\exp \,[\, 2\zeta(\omega_i)(z+\omega_i)\, ]\, .
 \end{array}
\end{equation}
The addition formulae for these functions are
\begin{eqnarray}
 \label{ad1} \wp(z_1+z_2)&=&
   \frac{1}{4}\left [\frac{\dot{\wp}(z_1)-\dot{\wp}(z_2)}{\wp(z_1)-\wp(z_2)}\right
   ]^2-\wp(z_1)-\wp(z_2)\, ,\\ [1ex]
 \label{ad2}\zeta (z_1+z_2)&=&\zeta (z_1)+\zeta
  (z_2)+\frac{1}{2}\frac{\dot{\wp}(z_1)-\dot{\wp}(z_2)}{\wp(z_1)-\wp(z_2)}\, .
\end{eqnarray}

 We will now derive the energy equation (\ref{en}).
 Using the relations (\ref{om})--(\ref{odd}) and the addition
formula (\ref{ad2}), and noting that
\begin{equation}
 \begin{array}{ll}
 \wp (z+2\omega_i)=\wp (z)\, ,& \wp (-z)=\wp (z)\, , \\ [1ex]
 \dot{\wp}(z+2\omega_i)=\dot{\wp}(z)\, , & \dot{\wp}(-z)=-\dot{\wp}(z)\, ,
 \end{array}
\end{equation}
 it is not very difficult to derive the following relations
 \begin{equation}\label{fin}
 \dot{\chi}_1^{\pm}=A_{\pm}\chi_1^{\pm}\, , \qquad
 \dot{\chi}_2^{\pm}=-A_{\mp}\chi_2^{\pm}\, ,
 \end{equation}
 where $\chi_i(z)$ and $A_{\pm}$ are given by (\ref{ls}) and
 (\ref{defa}). Inserting the expressions for $\dot{\chi}_i^{\pm}$
 from (\ref{fin}) into (\ref{def1a}),
 the energy equation (\ref{en}) will readily follow from
 (\ref{enp}).

\end{document}